\documentclass[aps,prb,reprint,groupedaddress,superscriptaddress,
showpacs,floatfix]{revtex4-1}

\usepackage{graphicx}
\usepackage{bm}
\usepackage{amsmath}
\usepackage{dcolumn}

\newcolumntype{.}{D{.}{.}{3}}  
\newcommand{\MCS}{$M$CrS$_2$}
\newcommand{\nn}{\nonumber}
\newcommand{\qv}{\ensuremath{\mathbf{q}}}
\newcommand{\LDAU}{LSDA+$U$}
\newcommand{\EF}{\ensuremath{E_F}}
\newcommand{\ttg}{\ensuremath{t_{2g}}}
\newcommand{\eg}{\ensuremath{e_{g}}}
\newcommand{\mb}{\ensuremath{\mu_{\text{B}}}}

\begin{document}

\title{Unusual magnetism of layered chromium sulfides $M$CrS$_2$
($M$\,=\,Li, Na, K, Ag, and Au)}

\author{A. V. Ushakov}
\affiliation{II. Physikalisches Institut, Universit\"at zu K\"oln,
Z\"ulpicherstra{\ss}e 77, D-50937 K\"oln, Germany}
\affiliation{Max-Planck-Institut f\"ur Festk\"orperforschung,
Heisenbergstra{\ss}e 1, D-70569 Stuttgart, Germany}
\affiliation{Institute for theoretical physics, Clausthal University of
Technology, Leibnizstra{\ss}e 10, D-38678 Clausthal Zellerfeld, Germany}

\author{D. A. Kukusta}
\affiliation{Max-Planck-Institut f\"ur Festk\"orperforschung,
Heisenbergstra{\ss}e 1, D-70569 Stuttgart, Germany}
\affiliation{Institute for Metal Physics, 36 Vernadskyi Bld., UA-03680 Kiev,
Ukraine}

\author{A. N. Yaresko}
\affiliation{Max-Planck-Institut f\"ur Festk\"orperforschung,
Heisenbergstra{\ss}e 1, D-70569 Stuttgart, Germany}

\author{D. I. Khomskii}
\affiliation{II. Physikalisches Institut, Universit\"at zu K\"oln,
Z\"ulpicherstra{\ss}e 77, D-50937 K\"oln, Germany}

\begin{abstract}

$M$CrS$_2$ compounds ($M$\,=\,Li, Na, K, Cu, Ag, and Au) with triangular Cr
layers show large variety of magnetic ground states ranging from
120$^{\circ}$ antiferromagnetic order of Cr spins in LiCrS$_2$ to double
stripes in AgCrS$_2$, helimagnetic order in NaCrS$_2$, and, finally,
ferromagnetic Cr layers in KCrS$_2$. On the base of \textit{ab-initio} band
structure calculations and an analysis of various contributions to exchange
interactions between Cr spins we explain this tendency as originating from a
competition between antiferromagnetic direct nearest-neighbor $d$--$d$
exchange and ferromagnetic superexchange via S\,$p$ states which leads to the
change of the sign of the nearest neighbor interaction depending on the
radius of a $M$ ion. It is shown that other important interactions are the
third-neighbor interaction in a layer and interlayer exchange. We suggest
that strong magneto-elastic coupling is most probably responsible for
multiferroic properties of at least one material of this family, namely,
AgCrS$_2$.

\end{abstract}


\maketitle

\section{\label{sec:introd}Introduction}

Frustrated magnetic systems attract now considerable
attention.\cite{Ramirez-94} Among them there are systems with very strong
geometric frustrations (e.g.\ kagome or pyrochlore systems), and also less
frustrated ones -- e.g.\ systems with triangular lattices. Triangular magnets
are overconstrained and most often they display one or the other type of
magnetic ordering. Nevertheless, frustrated nature of triangular layers
strongly influences their magnetic properties, often making them rather
unusual and very sensitive to small variations of the electronic and lattice
structure. \cite{Collins-97} Such materials also present definite practical
interest, e.g., as possible thermopower materials\cite{Terasaki-06} or new
multiferroics.\cite{Khomskii-09, Cheong-07}

The presence of orbital degeneracy may introduce special features in the
properties of triangular magnets, see e.g., Ref.~\onlinecite{Martin-11}. But
even without such degeneracy, as in materials containing half-filled
$d$-(sub)shells (Fe$^{3+}$ $d^5$; Cr$^{3+}$ $t_{2g}^3$), the properties of
such systems can be rather nontrivial.

Contrary to similar materials with oxygen instead of sulfur, $M$CrS$_2$
compounds are much less studied.  But it was recently shown that at least some
of them, such as AgCrS$_2$, show very interesting behavior: this particular
material belongs to a pyroelectric class, below the N\'{e}el temperature
$T_N$\,=\,50 K it develops a rather unusual double-stripe (DS) magnetic order
\cite{DMH+11} and also becomes multiferroic.\cite{Simon-10} Motivated by this
findings, and trying to understand the reasons for this unusual type of
magnetic ordering, apparently also relevant for the appearance of
ferroelectricity, we undertook a study of this and similar systems with the
$M$-ions Li, Na, K, Cu, Ag, Au. These systems, though in principle very
similar and all containing as the main building block the same CrS$_2$
triangular layers, show very different magnetic ordering: from the pure
nearest neighbor antiferromagnetism (120$^{\circ}$ structure) for
LiCrO$_2$~\cite{Loidl-07} with the smallest M-cation Li$^{+}$ and up to
ferromagnetic (FM) CrS$_2$ layers in KCrS$_2$~\cite{LE73} with the largest
$M$-ion K$^{+}$, with more complicated magnetic structures in the other
systems. Our \textit{ab-initio} and model calculations allow us to explain the
general tendency of magnetic ordering in this very rich class of compounds,
and this understanding may be helpful not only for these compounds, but also
for other magnetic systems with triangular layers.

\section{\label{sec:structure}Crystal and magnetic structure }

\begin{figure}[tbp!]
\includegraphics[width=0.16\textwidth]{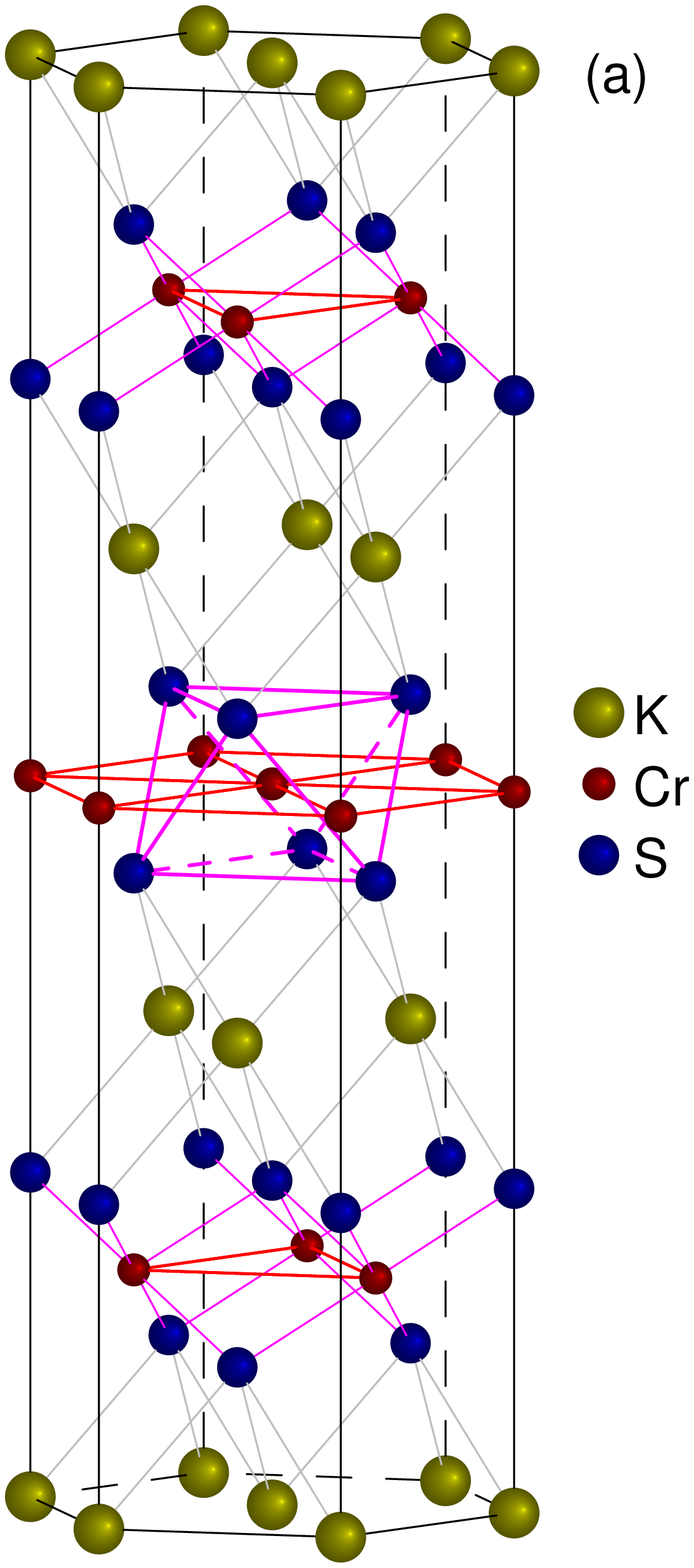}\ 
\includegraphics[width=0.15\textwidth]{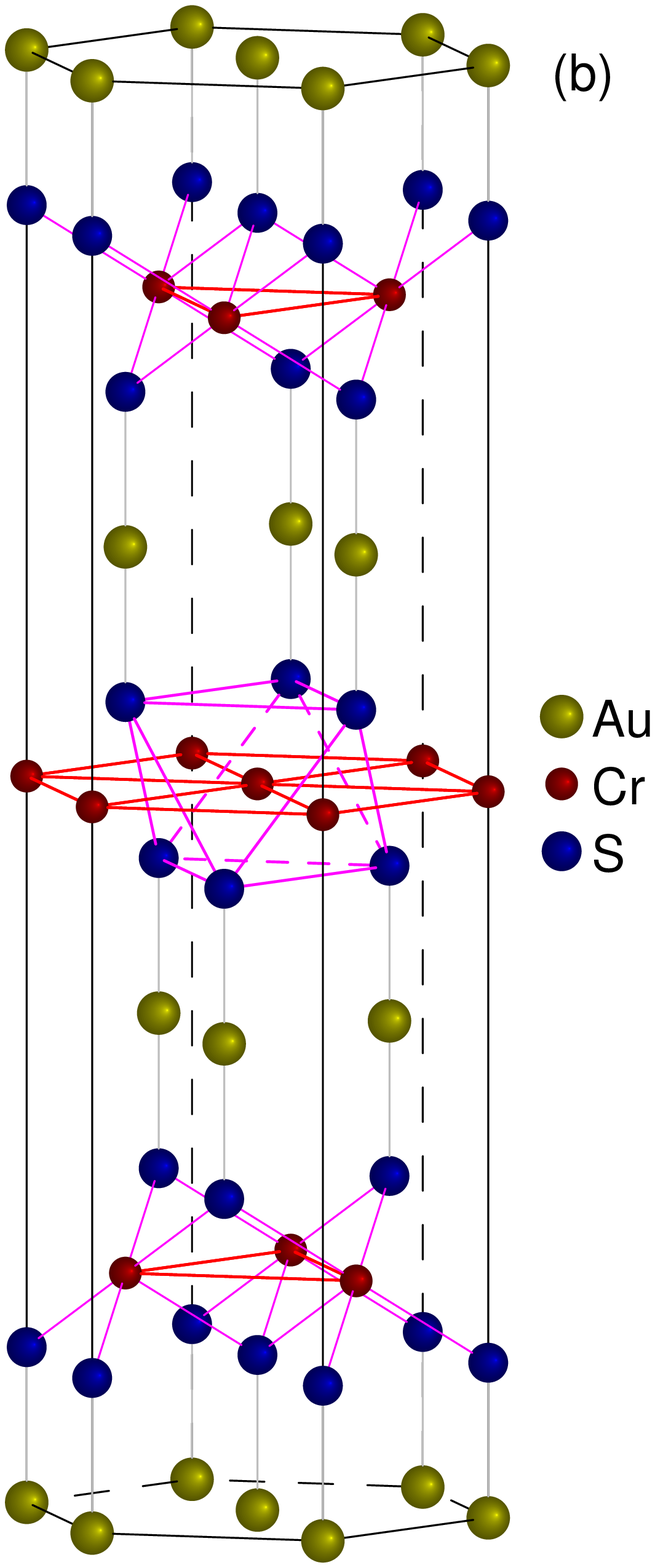}\ 
\includegraphics[width=0.16\textwidth]{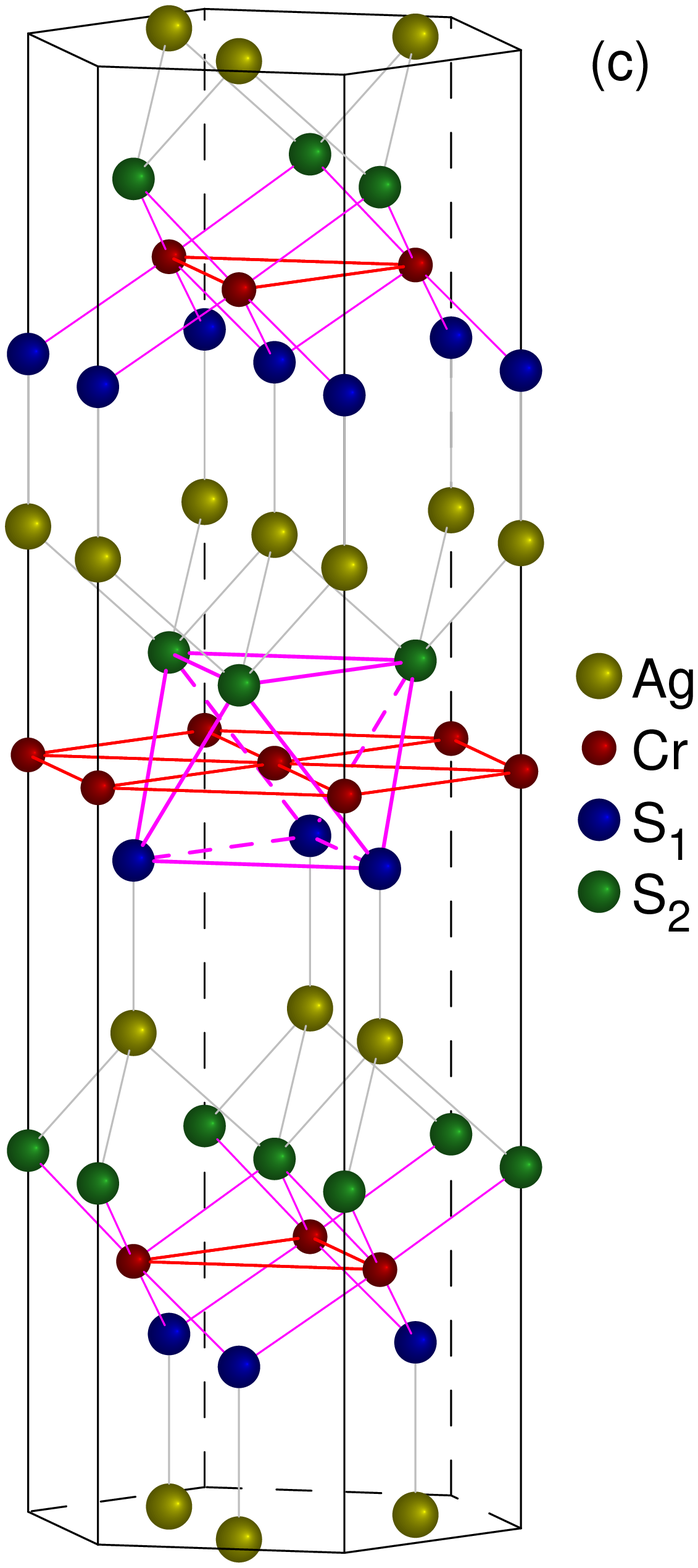}
\caption{\label{fig:str}(Color online) High temperature rhombohedral crystal
structures of KCrS$_2$ (a), AuCrS$_2$ (b), and AgCrS$_2$ (c). Also shown is a
distorted CrS$_6$ octahedron surrounding a Cr ion in the cell center.}
\end{figure}

The crystal structure of $M$CrS$_2$ series has been determined in Refs.\
\onlinecite{RS43,EWJ+73,LI71,DMH+11,CRY+11}. Cr atoms form a triangular
lattice within CrS$_2$ layers, and the latter are joined by $M$ atoms
[Fig.~\ref{fig:str}(a)]. Cr~atoms are located at the center of trigonally
distorted octahedra composed of sulfur ones. Each S atom is shared by three
different octahedra.  But the ``connection'' between layers is different in
different compounds. In compounds with alkali metals, Li, Na, and K are also
sitting in S$_6$ octahedra. One can visualize the structure of these compounds
as originating from the rock-salt structure of (actually hexagonal) CrS, in
which Cr and alkali ions are ordered in consecutive (1,1,1) planes, so that Cr
as well as Li, Na, or K are octahedrally coordinated by anions (the detailed
stacking of Cr, S and alkali layers maybe different).

At the same time, the structure of the systems $M$CrS$_2$ with $M$=Cu, Ag, Au
is different. In corresponding oxides the nonmagnetic ions Cu$^{1+}$ and
Au$^{1+}$ with the configuration $d^{10}$ are linearly coordinated. They are
located in the centers of oxygen dumbbells, i.e., are sandwiched between two
oxygens belonging to different MO$_2$) layers. The resulting structure is
that of delafossites. \cite{Pabst-46}

The structure of their sulfur analogues is more interesting: it is ``in
between'' that of, say, LiCrO$_2$ and AgCrO$_2$. Such $M^+$ ions are located
on top of a S$_3$ triangle of one, say, the lower CrS$_2$ layer, but are
connected by vertical bonds to one S$^{2-}$ ion of the next, upper layer
[Fig.~\ref{fig:str}(c)]. In effect Cu, Ag, and Au are in a ``tripod'' made of
four S ions, or in the distorted (elongated in c-direction) S$_4$ tetrahedron.
Metal ions in such S$_4$ tetrahedra are strongly shifted towards the upper,
apical S ion. All such tripods, or tetrahedra, are pointing in the same
direction, e.g., up, so that the resulting structure does not have an
inversion symmetry and is of a pyroelectric class. However this interesting
structural feature, though probably important for some properties of these
materials, seem to play minor role in magnetic properties of these compounds,
which mainly depend on interactions in CrS$_2$ layers.  Whereas most
structural studies of $M$CrS$_2$ with $M$=Cu and Ag give this structure with
$M$-ions in sulfur ``tripods'' and R3m symmetry,\cite{Martin-11} there are
also reports of a different crystal structure. Thus, in the recent paper
Ref.~\onlinecite{CRY+11} it is concluded that the symmetry of AuCrS$_2$ is
$R\overline{3}m$ or, maybe, $R3m$, and the actual structure is the delafossite
one with linearly coordinated Au$^+$ [Fig.~\ref{fig:str}(b)].

$M$CrS$_2$ compounds have diverse magnetic structures and a
broad set of physical properties. Being coupled antiferromagnetically
(AFM) between the layers, they exhibit quite different in-plane ordering
at low temperatures.

At high temperatures LiCrS$_2$ belongs to P$\overline{3}$m1 space symmetry
group. According to neutron scattering measurements the magnetic structure of
this compound below the N\'eel temperature $T_N$\,=\,55 K exhibits a
triangular spin arrangement (120$^{\circ}$ structure) within the triangular
planes, with adjacent planes being coupled
antiferromagnetically.\cite{Lafond-01, LI71} This structure is typical for
Heisenberg antiferromagnets with nearest neighbor coupling on a triangular
lattice.  The observed value of Cr$^{3+}$ spin magnetic moment equals 2.26\mb,
being considerably smaller than the expected value of 3\mb. The difference may
be presumably attributed to covalency effects, which can considerably alter
the distribution of the spin density around the Cr$^{3+}$ ion. Indeed, one can
expect such behavior, keeping in mind much smaller size of Li$^+$ ions and
respective reduction of unit cell volume.

KCrS$_2$ undergoes AFM transition at T$_N$\,=\,38\,K.\cite{LE73} the symmetry
group at high temperature is rhombohedral R$\overline{3}$m. The magnetic
structure, in contrast to LiCrS$_2$, consists of ferromagnetic layers
perpendicular to the $c$ axis, which are antiferromagnetically coupled to
adjacent ones. The paramagnetic Curie temperature of KCrS$_2$ is not low
($\theta_C$\,=\,112\,K) and indicates that the ferromagnetic interaction in
the planes is the dominant one. The observed value of the Cr$^{3+}$ spin
magnetic moment (3.04$\pm$0.05\mb) obtained by neutron scattering \cite{LE73}
is in good agreement with the expected value of 3\mb\ and with the value
obtained from the susceptibility measurements (3.1\mb). This can be
interpreted as an indication that in KCrS$_2$ covalency effects are relatively
weak.

In contrast to LiCrS$_2$ and KCrS$_2$, AgCrS$_2$ undergoes at
$T_N$\,=\,41.6\,K a first-order phase transition from a paramagnetic
rhombohedral $R3m$ structure to an antiferromagnetic monoclinic $Cm$
structure.\cite{DMH+11} Most interesting, the material was found to be
ferroelectric below $T_N$, i.e., it is a multiferroic system.\cite{Simon-10}
Note that this phenomenon differs from the eventual polarization of AgCrS$_2$
due to its pyroelectric crystal structure: this polarization appears only in a
magnetically-ordered state and lies in the $ab$-plane, not along
$c$-direction, as the eventual pyroelectric polarization due to the crystal
structure itself. In addition to being ferroelectric below $T_N$, the
low-temperature phase of AgCrS$_2$ exhibits an unconventional collinear
magnetic structure that can be described as double ferromagnetic stripes
coupled antiferromagnetically, with the magnetic moment of Cr$^{3+}$ aligned
along the $b$ axis within the anisotropic triangular plane. Ferroelectricity
below $T_N$ in AgCrS$_2$ is explained as a consequence of atomic displacements
at the magnetoelastically induced structural distortion, most probably driven
by the double-stripe magnetic structure itself. Thus, this system can be
classified as a type-II multiferroic.\cite{Khomskii-09, Khomskii-06,
Cheong-07}

Similarly to AgCrS$_2$, AuCrS$_2$ undergoes a first-order magnetic and
structural phase transition at $T_N$\,=\,47\,K from a paramagnetic
rhombohedral $R\overline{3}m$ to a monoclinic antiferromagnetic $C2/m$
structure. \cite{CRY+11} The simultaneous observation of magnetic and
structural transition both in AgCrS$_2$ and AuCrS$_2$ gives evidence of a
large magnetoelastic coupling in these systems. This coupling accounts for the
stability of the observed magnetic order, considering that the structural
distortions at the transition suppress the geometric frustration of the Cr
layers.  As we will show below, the peculiar antiferromagnetic structure
observed both in AgCrS$_2$ and AuCrS$_2$ is explained by the interplay of the
exchange due to direct $dd$ hopping and that via anions (sulfur) involving
nearest neighbor and further neighbor Cr-Cr interactions, as well as the
residual frustration in the triangular Cr planes.

In Table \ref{table:dist} we put different compounds in the order of
increasing Cr--Cr distance, which also corresponds to an increase of a
Cr--S--Cr bond angle since average Cr--S distances vary much less than the
Cr--Cr ones. One immediately notices a definite correlation between the
crystal structure and magnetic order: with increasing Cr--Cr distance and
Cr--S--Cr angle the magnetic structure changes from the 120$^{\circ}$ AFM
structure in LiCrS$_2$ with the smallest Li$^{+}$ ion and the shortest Cr--Cr
distance to ferromagnetic layers in KCrS$_2$ with the largest K$^{+}$ ion and
the longest Cr--Cr distance. The crossover between these limiting cases occurs
via incommensurate magnetic phases in CuCrS$_2$ and NaCrS$_2$ and the
double-stripe structure in AuCrS$_2$ and AgCrS$_2$. It is this correlation
between crystal and magnetic structure, which is the main topic of our
study. We approach this problem by performing \textit{ab initio} calculations,
in which we obtain the electronic structure of the {\MCS} compounds, as well
as the values of relevant exchange constants. We then analyze the observed
general trends in a superexchange model, discussing different relevant, often
competing contributions to the total exchange.

\begin{table}[tbh]
  \caption{\label{table:dist}Cr--Cr ($d_{\text{Cr-Cr}}$) and Cr--S
  ($d_{\text{Cr-S}}$) interatomic  distances  (in \AA) as well as Cr-S-Cr bond
  angles $\theta$ (in degrees) for the high-temperature $M$CrS$_2$
  structures. For $M$=Cu and Ag only averaged $d_{\text{Cr-S}}$ is shown, while 
  $\theta$ is given for two inequivalent S ions.}
\begin{ruledtabular}
\begin{tabular}{c ccc c}
 $M$ & $d_{\text{Cr-Cr}}$ & $d_{\text{Cr-S}}$ & $\theta$ & magnetic structure \\
\hline
  Li & 3.4515 & 2.4063 & 91.7 & AFM 120$^{\circ}$  \\	
  Cu & 3.4728 & 2.4036 & 90.6, 94.6 & spiral ordering \\
  Au & 3.4826 & 2.3862 & 93.7 & AFM double stripes \\
  Ag & 3.4979 & 2.4085 & 92.2, 94.1 & AFM double stripes \\
  Na & 3.5561 & 2.4249 & 94.3 & spiral ordering \\
  K  & 3.6010 & 2.4123 & 96.6 & FM in plane \\
\end{tabular}
\end{ruledtabular}
\end{table}

\section{\label{sec:details}Computation details}

Band structure calculations were performed using the linear muffin-tin
orbitals (LMTO) method~\cite{And75} as implemented in the PY-LMTO computer
code.~\cite{PYAunp} We used the Perdew-Wang~\cite{PW92} parameterization for
the exchange-correlation potential in the local spin-density approximation
(LSDA). Brillouin zone integrations were performed using the improved
tetrahedron method.~\cite{BJA94}

When the spin-orbit coupling is not taken into account, the use of the
generalized Bloch theorem \cite{San91a} makes possible self-consistent
calculations of the band structure and the total energy $E(\qv)$ for
spin-spiral structures with an arbitrary wave vector \qv\ as described in
details in Refs.\ \onlinecite{YPHR02,Yar08}.  In these calculations the
magnetization direction in an atomic sphere centered at
$\mathbf{t}+\mathbf{R}$, where $\mathbf{t}$ specifies its position in a unit
cell and $\mathbf{R}$ is a translation vector, is defined by two polar angles
$\theta$ and $\phi=\qv \cdot \mathbf{R} + \phi_{0}$. In the present work we
considered only planar spin spirals with all $\theta=\pi/2$. The phase
$\phi_{\mathrm{Cr}}$ inside spheres surrounding Cr ions was fixed by requiring
$\phi_{\mathrm{Cr}}=\qv \cdot \mathbf{t}_{\mathrm{Cr}}$, whereas for all other
spheres it was determined selfconsistently by diagonalizing the corresponding
spin-density matrix.

This general approach allows us to treat on the same footing not only
collinear, e.g., ferromagnetic or stripe, or non-collinear, e.g.,
120$^{\circ}$ AFM, \emph{commensurate} magnetic structures, but also perform
calculations for \emph{incommensurate} helical structures. The only
restriction is that it should be possible to describe the magnetic structure
by a single wave vector \qv. After the \qv\ dependence of the total energy has
been calculated, effective exchange interactions between Cr spins can be
obtained by mapping $E(\qv)$ onto a relevant Heisenberg-like model.

The magneto-crystalline anisotropy was estimated by using the force theorem,
\cite{MA80} i.e., by comparing band energies obtained for selected collinear
spin structures from spin-polarized relativistic calculations with the
magnetization parallel to different crystallographic axes. Spin-orbit coupling
in these calculations was included into the LMTO Hamiltonian at the
variational step.\cite{PFK80}

In order to study the effect of relatively strong electronic correlations in
the Cr $d$ shell on the band structure and magnetic interaction in the {\MCS}
compounds, for some of them we also calculated $E(\qv)$ using the rotationally
invariant \LDAU\ method.\cite{Anisimov-97} For the double counting term the
so-called atomic limit was used.\cite{CS94} Other details on the
implementation of the \LDAU\ method in the PY-LMTO code are given in Ref.\
\onlinecite{YAF03}. Calculations were performed for the Hund's exchange
coupling parameter $J=0.9$ eV and the on-site Coulomb repulsion $U=1.9$, 2.9,
and 3.9 eV, which gives 1, 2, and 3 eV for $U_{\text{eff}}=U-J$.

\section{\label{sec:results}Results and discussion}

\subsection{\label{sec:Ag_BNS}Band structure and energies of different
magnetic structures}

Our band-structure calculations demonstrate that all atoms in $M$CrS$_2$
compounds exhibit their valences corresponding to the stoichiometry of the
compound, i.e., the atomic charges correspond to $M^+$, Cr$^{3+}$, and
S$^{2-}$. The $s$~orbitals of $M^+$ are empty, whereas the $p$~orbital of
S$^{2-}$ are fully occupied. Since Cr atom is triply ionized, there are three
$d$ electrons localized on a Cr$^{3+}$ ion.

The octahedral crystal field at the Cr site causes the $d$ orbitals to split
into a triplet $t_{2g}$ ($xy$, $xz$, $yz$) and a doublet $e_g$ ($3z^2-r^2$,
$x^2-y^2$), with the energy of the $t_{2g}$ orbitals being lower than that of
the $e_g$ states. Since there are three $d$~electrons localized on a Cr site,
in spin-restricted band structure calculations the $t_{2g}$ states are
half-filled, whereas the $e_g$ levels are empty. In spin-polarized
calculations the spin-up $t_{2g}$ states are occupied, and spin-down $t_{2g}$
are empty.

\begin{figure}[tbp!]
\includegraphics[width=0.46\textwidth]{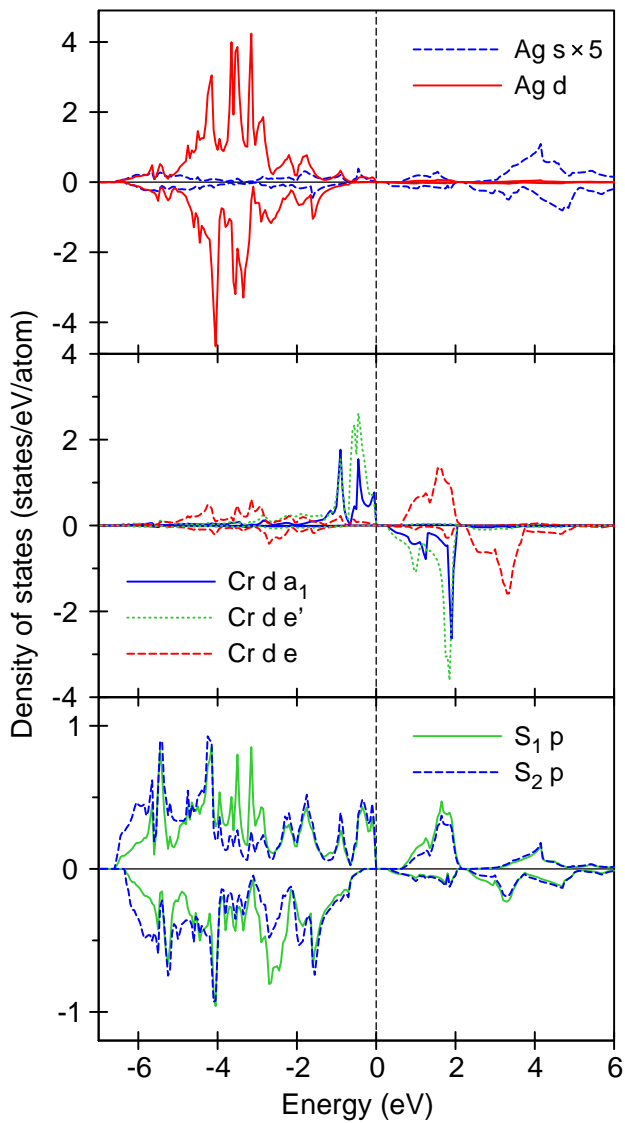}
\caption{\label{fig:Ag_DOS}(Color online) Partial densities of states in
AgCrS$_2$ with the FM alignment of Cr moments. Energies are given
relative to the Fermi level \EF.} 
\end{figure}

The cubic component of the crystal field at the Cr site is strong enough for
the $t_{2g}$ and $e_g$ orbitals to form two non-overlapping sub-bands
separated by an energy gap of about $0.5$\,eV.  Additional trigonal distortion
along the $c$-axis lifts the degeneracy of the $t_{2g}$ levels and splits them
into a singlet $a_{1g}$ and a doublet $e_g'$ ($a_{1}$ and $e'$ in the
compounds with $R3m$ symmetry, e.g., AgCrS$_2$), which are linear combinations
of the $t_{2g}$ orbitals. Three Cr$^{3+}$\,$d$ electrons occupy spin-up
$a_{1g}$ and $e_g'$ orbitals.

The electronic structures and density of states (DOS) of compounds in
$M$CrS$_2$ series are similar, so to get details specific for current
calculations, we consider as an example the DOS’es obtained for AgCrS$_2$ in
ferromagnetic spin-polarized LSDA calculations (see Fig.~\ref{fig:Ag_DOS}).

The occupied part of the valence band can be subdivided into several regions.
For all $M$-ions their valence $s$ states are empty and $d$ states (if they
exist) are totally occupied. These valence $s$ and $d$ states do not
contribute to the electronic density close to the Fermi-energy {\EF}. In
AgCrS$_2$ the Ag\,4$d$ states appear between $-$6\,eV and $-$1.5\,eV.

The occupied S$^{2-}$\,3$p$ states form the broad band with the width of 6\,eV
between $-$6.5\,eV and $-$0.3\,eV, being strongly hybridized both with
Ag\,4$d$ and Cr\,3$d$ states. As will be discussed later, this hybridization
between Cr\,3$d$ and S\,$3p$ states is responsible for superexchange along
Cr--S--Cr and Cr--S--S--Cr paths. According to our band structure calculations
these materials are insulating even in the ferromagnetic state and even
without including Hubbard's U. For instance for AgCrS$_2$ the LSDA gives the
energy gap of 0.55 eV. That is, due to their specific electronic structure --
half-filled $t_{2g}$ subshell and empty $e_g$ states of Cr$^{3+}$ -- they
would be band insulators (in a magnetically-ordered state).  When electronic
correlations are accounted for in {\LDAU} calculations, the occupied
majority-spin \ttg\ states are shifted by $U_{\text{eff}}/2$ to lower
energies, whereas the unoccupied minority-spin \ttg\ and all \eg\ states move
$\sim U_{\text{eff}}/2$ to higher energies which increases the values of the
gaps.

Magnetic properties and the electronic structure of $M$CrS$_2$ compounds are
closely related to the occupancy of the Cr\,3$d$ states, which are spread over
wide energy interval from $-$6\,eV to 4\,eV and form two non-overlapping
subbands separated by energy gap. Cr $e_g$ and $t_{2g}$ orbitals form
$pd\sigma$- and $pd\pi$-bonds with sulfur $p$~orbitals, respectively.  The
hybridization between occupied Cr spin-up $t_{2g}$ and S\,$p$-states at
$-$1\,eV and $-$0.5\,eV is clearly observed. Being rather small below {\EF},
the hybridization between Cr\,$d$ and S\,$p$ above {\EF} is larger for $e_g$
states and is well pronounced for spin-up DOS’es.

Our calculations prove the clearly insulating nature of these materials. The
exchange splitting $\Delta_{\rm ex}$\,$\sim$\,2\,eV is prominent for the
Cr\,3$d$ bands in the whole $M$CrS$_2$ series where only spin-up Cr\,$a_{1g}$
and $e_g'$ orbitals are filled.

The calculated values of the Cr spin magnetic moment are close to 3\mb\ for
all compounds in the series. Calculations for spin spirals showed that the Cr
moment depends only weakly on the wave vector of a spiral, i.e., on the kind
of magnetic order. In LiCrS$_2$, for instance, the moment varies from 2.74\mb\
for the 120$^{\circ}$ AFM structure to 2.98\mb\ for the FM one. This also
confirms the localized character of the Cr moments and suggests that magnetic
interactions between them can be described by the Heisenberg model.

Damay \textit{et al.}\ in Ref.~\onlinecite{DMH+11} analyzed dynamic
correlations and found a small spin gap at very low energies as $\mathbf{q}\to
0$ that has been attributed to the weak magnetic anisotropy; i.e., we conclude
that the Cr spins in $M$CrS$_2$ are relatively isotropic and can be described
by the Heisenberg model. The localized character of Cr$^{3+}$ spin magnetic
moments is confirmed in our calculations by the fact that Cr spin-up $a_{1g}$
and $e_g'$ states are fully occupied, localized on the Cr$^{3+}$ site and
separated from empty states by an energy gap.

\begin{figure}[tbp!]
\includegraphics[width=0.45\textwidth]{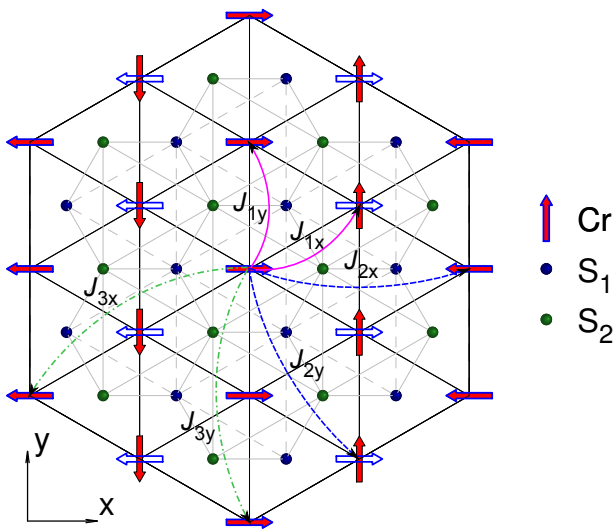}
\caption{\label{fig:90_ds}(Color online) Representation of double stripe (blue
arrows) and 90$^\circ$ (red arrows) magnetic structures within the Cr plane.
One underlying (S$_1$) and one overlying (S$_2$) sulfur layers are shown as
well. The low-temperature intraplane exchanges between the first ($J_{1x}$,
$J_{1y}$), the second ($J_{2x}$, $J_{2y}$), and the third ($J_{3x}$, $J_{3y}$)
neighbors are shown as curved lines with arrows.  In the high-temperature
phase $J_{1x}$\,=\,$J_{1y}$\,=\,$J_1$, $J_{2x}$\,=\,$J_{2y}$\,=\,$J_2$,
$J_{3x}$\,=\,$J_{3y}$\,=\,$J_3$.  }
\end{figure}

The applicability of the Heisenberg model allows us to investigate the wide
range of Cr spin moment configurations within the single approach using the
Heisenberg Hamiltonian in the form
\begin{eqnarray}
H={1\over 2}\sum_{i\ne j} J_{ij}\mathbf{S}_i\mathbf{S}_j .
\label{Ham}
\end{eqnarray}

Everywhere below we work in the orthogonal coordinates, choosing $y$-axis
along one of the directions between Cr-Cr nearest neighbors in the $ab$-plane,
and the $x$-axis is chosen perpendicular to it. i.e., it points from one Cr to
its second neighbor, see Fig.~\ref{fig:90_ds}. Thus, in our notation the
\qv-vectors of magnetic superstructures are given in these coordinates, not in
the standard vectors of corresponding reciprocal lattices. We measure the
in-plane components of a \qv-vector in the units of $2\pi/a$ and the
out-of-plane component in $2\pi/c$.

In case of an arbitrary wave vector \qv\,=\,($q_x$, $q_y$, $q_z$) the
Heisenberg magnetic energy in these coordinates is
\begin{eqnarray}
E(\mathbf{q})=\epsilon_1(\mathbf{q})+\epsilon_2(\mathbf{q})
+\epsilon_3(\mathbf{q})+\epsilon_z(\mathbf{q})
\label{E_q}
\end{eqnarray}
where $\epsilon_i(\mathbf{q})$ are contribution proportional to the exchange
coupling constants $J_i$ between $i$-th Cr neighbors within the triangular
plane (see Fig.\ \ref{fig:90_ds}). For the undistorted high-temperature (HT)
rhombohedral structures
\begin{eqnarray}
\epsilon_1(\mathbf{q})&=&J_{1}\left[2\cos(\sqrt{3}q_xa/2)\cos(q_ya/2) 
\right. \nn \\
&&  \left. +\cos(q_ya)\right],
\label{E_q1}
\end{eqnarray}
\begin{eqnarray}
\epsilon_2(\mathbf{q})&=&J_{2}\left[\cos(q_xa\sqrt{3}) \right. \nn \\
&&\left. +2\cos(\sqrt{3}q_xa/2)\cos(3q_ya/2) \right],
\label{E_q2}
\end{eqnarray}
\begin{eqnarray}
\epsilon_3(\mathbf{q})&=&J_{3}\left[2\cos(q_xa\sqrt{3})\cos(q_ya)+\cos(2q_ya)
  \right].
\label{E_q3}
\end{eqnarray}

An expression for interlayer coupling, $\epsilon_z(\mathbf{q})$, is
particularly simple for LiCrS$_2$:
\begin{eqnarray}
\epsilon_z(\mathbf{q}) &=& J_{z} \cos(q_zc) \, .
\label{E_qzli}
\end{eqnarray}
In other compounds with the $abc$ stacking of Cr layers Cr neighbors in
adjacent planes sit above and below the centers of triangles, i.e., above
S$_1$ and below S$_2$ positions in Fig.\ \ref{fig:90_ds}, and
$\epsilon_z(\mathbf{q})$ becomes
\begin{eqnarray}
\epsilon_z(\mathbf{q}) &=& J_{z}\left[2\cos(q_z c/3-q_xa/(2\sqrt{3}))
\cos(q_ya/2) \right.
\nonumber\\
&& \left. +\cos(q_zc/3+q_xa/\sqrt{3})\right].
\label{E_qz}
\end{eqnarray}
When $J_z$ is sufficiently strong it may affect in-plane magnetic order.

As sketched in Fig.~\ref{fig:90_ds}, in the monoclinic low-temperature (LT)
phases of AgCrS$_2$ and AuCrS$_2$ exchange interactions $J_{nx}$ and $J_{ny}$
between $n$-th neighbors along $x$ and $y$ directions are no longer equal and
the expressions (\ref{E_q1})--(\ref{E_q3}) should be modified accordingly.
For instance, the energy of the nearest neighbor interaction becomes
\begin{eqnarray}
\epsilon_1(\mathbf{q})&=&2J_{1x}\cos(q_x x_{1x})\cos(q_y y_{1x}) \nn \\
  && +J_{1y}\cos(q_y y_{1y}),
\label{E_q1lt}
\end{eqnarray}
where a vector $\mathbf{r}_{1x/y}$ ($x_{1x/y}$,$y_{1x/y}$,0) connects a Cr
site with its nearest neighbors along $x$ and $y$ directions.

\begin{figure}[tbp!]
\includegraphics[width=0.46\textwidth]{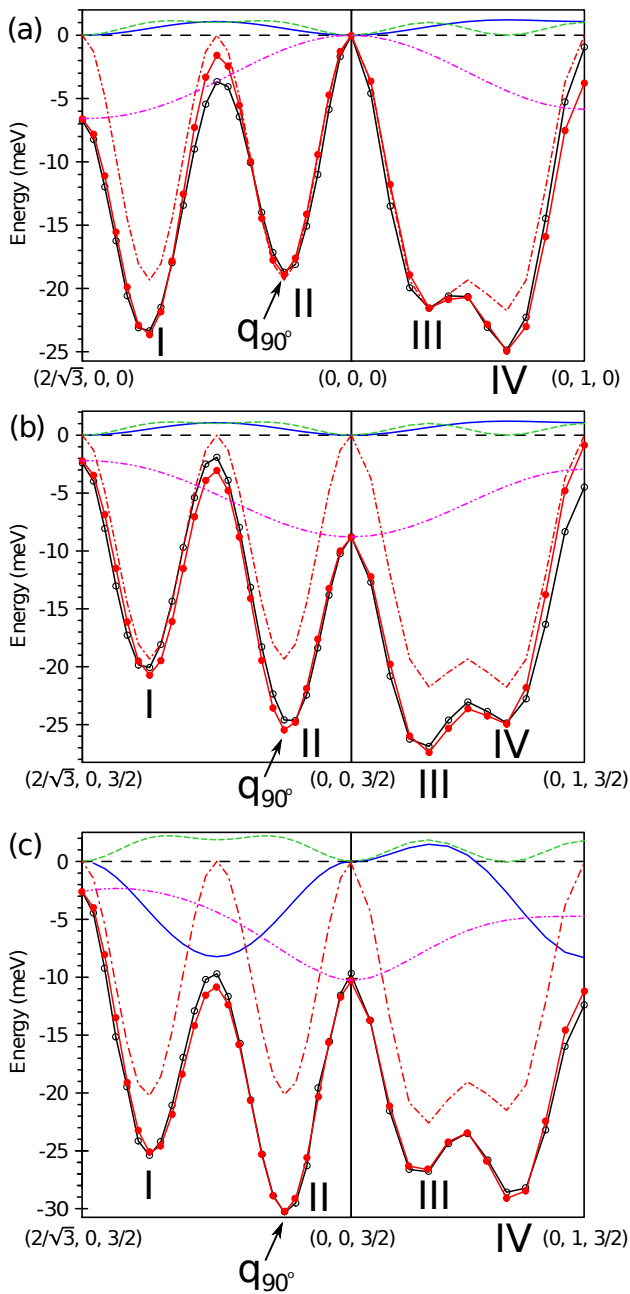}
\caption{\label{fig:Emq}(Color online) Calculated (open black circles) and
fitted (full red circles), using the least-squares method in Heisenberg model,
profiles $E(\mathbf{q})$ of magnetic energies for AgCrS$_2$ for two sets of
wave vectors, see the text.  Partial contributions in magnetic energy,
calculated according to (\ref{E_q1})--(\ref{E_qz}), are shown as well. The
$\epsilon_1(\mathbf{q})$, $\epsilon_2(\mathbf{q})$, $\epsilon_3(\mathbf{q})$,
and $\epsilon_z(\mathbf{q})$ profiles are presented by solid blue, dashed
green, dashed with one dot red, and dashed with two dots magenta lines.  The
dispersion curves shown in panels (a) and (b) are calculated for the
high-temperature phase with ferro (a) and antiferro (b) interlayer ordering,
whereas those in panel (c) are calculated for the low-temperature structure
and antiferro interlayer ordering. The $90^{\circ}$ structure, by which we
model the double-stripe magnetic structure observed in AgCrS$_2$ and
AuCrS$_2$, is marked by arrow.}
\end{figure}

In order to estimate the effective exchange parameters $J_i$ and $J_z$ we
first carried out \textit{ab-initio} calculations for a number of \qv-vectors
lying in $q_z$=0 [Fig.~\ref{fig:Emq}(a)] and $q_z$=3/2 [Fig.~\ref{fig:Emq}(b)]
planes. The latter value of $q_z$ results in 180$^{\circ}$ rotation of Cr
spins in adjacent layers. We then fitted the \qv-dependence of the calculated
total energy $E(\qv)$ (open black circles in Fig.~\ref{fig:Emq}) by the
Heisenberg model given by (\ref{E_q})--(\ref{E_qz}) using a least-square fit
with four ($J_1$, $J_2$, $J_3$, and $J_z$) and seven ($J_{1x,1y}$,
$J_{2x,2y}$, $J_{3x,3y}$, $J_{z}$) exchange parameters for the HT and LT
phases, respectively. The results of such a fit for the most interesting
system AgCrS$_2$, which has the unusual double-stripe magnetic structure and
becomes multiferroic below $T_N$, are shown in Fig.~\ref{fig:Emq} by filled
red circles. A good agreement between the results of the LSDA total energy
calculation and of the fit proves the possibility to describe the magnetic
properties of these compounds by the Heisenberg model which includes the
exchange coupling constants between first, second, and third neighbors, plus
interlayer exchange constant $J_z$. From these calculations we can extract the
values of the exchange constants for different materials, and by comparing the
energies of different states we can determine which state would be the ground
state for one or the other system.

\subsection{\label{sec:magstr}Magnetic structures}

Let us discuss the nature of different competing states i.e., different minima
of $E(\mathbf{q})$ in Fig.~\ref{fig:Emq}.  The 120$^{\circ}$ AFM spin
structure observed in LiCrS$_2$ (Ref.~\onlinecite{LI71}) is realized by a spin
spiral with \qv=(0,2/3) (in $2\pi/a$ units) which gives the minimum IV in
Fig.~\ref{fig:Emq}. The local minimum III corresponds to 120$^{\circ}$ AFM
order in the sublattice of 3-rd Cr neighbors.

The FM in-plane structure observed in KCrS$_2$ in Ref.~\onlinecite{LE73}
corresponds to $\qv=0$. In contrast, the magnetic energy of AgCrS$_2$ (Fig.\
\ref{fig:Emq}) has a maximum instead of a minimum at this \qv\ which agrees
with the fact that for this system the dominant exchange interactions are
antiferromagnetic.

The double-stripe spin structure observed in AgCrS$_2$ cannot be represented
as a single \qv\ spiral if the rhombohedral unit cell of the HT $R3m$
structure is used. However, it can be easily verified that the Heisenberg
energy of the DS structure is exactly equal to the energy of a spin spiral
with $\qv_{90^\circ}$=($\sqrt{3}/6$,0), shown by red arrows in Fig.\
\ref{fig:90_ds}, in which spins of each $i$-th Cr chain running along the $y$
direction turn by 90$^\circ$ with respect to the previous ($i-1$) one.

Indeed, let us consider the interaction of a Cr spin from some ($i$=0) chain
with the rest of the Cr plane. In the DS structure the spin directions in odd
chains to the left ($-2|i|+1$) and to the right ($2|i|+1$) are opposite and
their contributions to the magnetic energy $J\mathbf{S}_{0}\cdot
\mathbf{S}_{2|i|+1}$=$-J\mathbf{S}_{0}\cdot \mathbf{S}_{-2|i|+1}$ cancel each
other. In the 90$^{\circ}$ structure odd chains do not contribute to the
magnetic energy because of the orthogonality of Cr spins in odd and even
chains ($J\mathbf{S}_{0}\cdot \mathbf{S}_{2i+1}=0$). Consequently, the
magnetic energy is determined by the interaction of $\mathbf{S}_{0}$ with
$\mathbf{S}_{2i}$ from the even chains which are exactly the same in both spin
structures. Here we assume that the exchange coupling constants of
$\mathbf{S}_{0}$ with spins from chains to the left $\mathbf{S}_{-|i|}$ and to
the right $\mathbf{S}_{|i|}$ are equal. The couplings between $n$-th neighbor
lying in the same ($J_{ny}$) and different ($J_{nx}$) chains need not to be
equal so that the degeneracy of the DS and 90$^{\circ}$ structures holds also
for distorted Cr layers of the monoclinic LT phase of AgCrS$_2$.

LSDA supercell calculations performed for the DS and 90$^{\circ}$ structures
also gave the total energies which are equal within the numerical accuracy;
with their energy difference being less than 1 meV per Cr ion.  Because of the
degeneracy of the two spin structures the energy of the DS structure can be
calculated within the same spin-spiral approach as the energies of other
competing magnetic states. The corresponding energy minimum is marked as II in
Fig.\ \ref{fig:Emq}.

Experimentally, however, these two structures would lead to somewhat
different features of neutron scattering spectra although the positions of
magnetic Bragg peaks are the same. The authors of~[\onlinecite{DMH+11}]
concluded that the DS structure better fits the experimental data than the
90$^\circ$ structure.

The same 90$^{\circ}$ structure within a Cr plane is also realized at
\qv=($3\sqrt{3}/6$,0) corresponding to the minimum I in Fig.\
\ref{fig:Emq}. However, because of the rather strong interlayer coupling given
by Eq.\ (\ref{E_qz}) the energies at the minima I and II are not
equal. Finally, the maximum at \qv=($2\sqrt{3}/6$,0) between these two minima
corresponds to single-stripe magnetic order, in which FM Cr chains running
along $y$ are ordered antiferromagnetically.

Comparing the energies of different states in Fig.~\ref{fig:Emq}, we can make
several conclusions. First of all, we see that if the ordering between planes
would be ferromagnetic and without extra lattice distortion,
Fig.~\ref{fig:Emq}(a), the absolute minimum for the parameters calculated for
AgCrS$_2$ would correspond to the simple 120$^{\circ}$ AF structure, i.e., the
state IV in Fig.~\ref{fig:Emq}. Such in-plane ordering is indeed realized in
LiCrS$_2$, but for real AgCrS$_2$ the observed ordering is different and
corresponds to the double stripe structure.

When we change the interlayer ordering, making it antiferromagnetic, the
situation already changes: the 120$^{\circ}$ state (state IV) is destabilized,
and another state, III, becomes the absolute minimum, Fig.~\ref{fig:Emq}(b). We
also notice that AFM interlayer ordering strongly lowers the energy of the
90$^{\circ}$ structure (state II in Fig.~\ref{fig:Emq}(b)) which, as discussed
above, is degenerate with the DS one, so that this state starts to compete
with the state III.
And when we include the lattice distortion present in AgCrS$_2$ in the LT
phase, Fig.~\ref{fig:Emq}(c), the double stripe state II becomes the absolute
minimum. Thus, we see that for the lattice corresponding to the real LT
structure of AgCrS$_2$ the double-stripe magnetic ordering with the
antiferromagnetic coupling between layers is indeed the ground state in our
calculations. We also see that several factors are important for the
stabilization of such DS structure: besides particular ratio of different
exchange constants, see below, also particular $3D$ interlayer ordering and
lattice distortion, accompanying magnetic ordering, are all important for
making double-stripe structure.

But we also see from Fig.~\ref{fig:Emq} that there exist, especially in the HT
lattice, other magnetic states competing with the double-stripe one. Thus, one
can predict that the magnetic fluctuations above $T_N$, which could be probed
e.g.\ by inelastic neutron scattering, could be most pronounced not at the
wave vector corresponding to the double-stripe ground state structure, but at
other values of {\bf q}, for instance those corresponding to the solutions III
and IV in Fig.~\ref{fig:Emq}.

Yet one more conclusion which we can extract from Fig.~\ref{fig:Emq} is that,
at least in AgCrS$_2$, the spin-lattice (magnetostriction) coupling is very
important in these systems: only when we included the lattice distortion,
occurring in AgCrS$_2$ below $T_N$, did we obtain the real double-stripe
structure as a ground state.

In addition to the nonrelativistic calculations discussed above, we have also
studied the magneto-crystalline anisotropy in $M$CrS$_2$ by accounting for
spin-orbit coupling in calculations for the FM spin structure with the
magnetization directed along different crystallographic axes. It turns out
that Cr atoms form an easy-plane magnet, which is consistent with the
experimental results: \cite{RS43,EWJ+73,LI71,DMH+11,CRY+11} the spin-orbit
coupling rotates all Cr spin magnetic moments into the ab-plane even in
high-temperature phase, but does not affect the magnetoelastic in-plane
coupling and low-temperature lattice distortion.

\subsection{\label{sec:exch_const}Exchange constants}

The LSDA exchange parameters estimated for the HT structure of all six \MCS\
compounds by fitting corresponding $E(\qv)$ using the Heisenberg model defined
by (\ref{E_q})--(\ref{E_qz}) are presented in Table~\ref{table:HTJs_LSDA}. We
first do not consider the LT phases of AgCrS$_2$ and AuCrS$_2$, because we
want to concentrate on general trends observed in this whole class of
materials. Detailed results for the LT phases will be presented below. The
dependence of the exchange constants on $U$ in LSDA+$U$ calculations will be
discussed in Sec.~\ref{sec:ldau}.

From the Table~\ref{table:HTJs_LSDA} we see that, with the exception of
AuCrS$_2$ which deviates from the general trend and will be discussed below,
the variation of the nearest-neighbor exchange $J_1$ in the series $M$ = Li,
Cu, Ag, Na, K clearly correlates with the corresponding structural parameters
from Table~\ref{table:dist}.  With the increase of the size of $M$ ion and of
the Cr--Cr distance, $J_1$ changes form strongly antiferromagnetic in
LiCrS$_2$, with the smallest Li and shortest $d_{\text{Cr-Cr}}$, to strongly
ferromagnetic in KCrS$_2$, with the largest K and longest $d_{\text{Cr-Cr}}$,
and becomes very small in the Cu and Ag compounds with intermediate Cr--Cr
distances.

We also notice that in all the compounds the third neighbor exchange $J_3$ is
antiferromagnetic and rather strong. On the other hand, the second neighbor
exchange $J_2$ is weak and can in most cases be neglected.  Apparently it is
an interplay of the nn exchange $J_1$ and the third neighbor exchange $J_3$
which is primarily responsible for the stabilization of one or the other spin
structure in the $M$CrS$_2$ series.

\begin{table}[tbh]
\caption{\label{table:HTJs_LSDA}Different exchange coupling constants (in
meV) in the high-temperature phase of $M$CrS$_2$ calculated in LSDA.}
\begin{ruledtabular}
\begin{tabular}{c.....}
  $M$ & \multicolumn{1}{c}{$J_1$} & \multicolumn{1}{c}{$J_2$}
& \multicolumn{1}{c}{$J_3$} & \multicolumn{1}{c}{$J_z$} &
\multicolumn{1}{c}{$J_1/J_3$} \\
  \hline
   Li &  5.17 &  0.46 & 2.73 & 0.93 & 1.9 \\ 
   Cu &  0.16 &  0.03 & 1.51 & 0.82 & 0.1 \\
   Au &  7.41 &  1.63 & 5.93 & 2.93 & 1.3 \\
   Ag & -0.14 & -0.13 & 2.45 & 0.74 & -0.2 \\
   Na & -4.06 &  0.23 & 2.49 & 0.09 & -1.6 \\
   K  & -5.45 &  0.19 & 2.11 & 0.05 & -2.6 \\
\end{tabular}
\end{ruledtabular}
\end{table}

Taking these considerations into account it seems reasonable to apply the
$J_1$--$J_3$ model to investigate magnetic ordering in $M$CrS$_2$. It is well
known that the simple $J_1$ model with antiferromagnetic $J_1$\,$>$\,0 (see
e.g., Ref.~\onlinecite{CZ09}) gives noncollinear magnetic ground states with
\qv\,=\,(0, 2/3) and angles of 120$^\circ$ between spin magnetic moments.  In
the $J_1$--$J_3$ model the magnetic energy equals
$E_{1,3}(\mathbf{q})$\,=\,$\epsilon_1(\mathbf{q})+\epsilon_3(\mathbf{q})$.  A
simple analysis shows that for positive $J_1$ and $J_3$ the wave vector
$\qv_{\text{IV}}$\,=\,(0, 2/3) is still the global minimum with the energy of
$E=-3/2 (J_1+J_3)$.
Here we consider only extrema at wave vectors lying on $x$ and $y$ axes. Other
symmetrically equivalent extrema can be obtained by applying $\pm 2\pi/3$
rotations to corresponding \qv. The numbering of the minima corresponds to the
notations in Fig.\ \ref{fig:Emq}.

For $|J_1|\!<\!4J_3$ a local minimum appears at $\qv_{\text{II}}\!=\!(q_x,0)$
on the $x$ axis, with $q_x$ defined by $\cos(\sqrt{3} \pi q_x)\!=\!-J_1/4J_3$.
When $J_1\!<\!J_3/2$ another minimum $\qv_{\text{III}}$ appears also on the
$y$ axis which becomes the global minimum for FM $J_1\!<\!0$. If $J_1\!=\!0$,
$\qv_{\text{III}}$\,=\,(0,1/3) corresponds to 120$^\circ$ order of 3-rd
neighbor spins. As the strength of FM $J_1$ increases, both $\qv_{\text{II}}$
and $\qv_{\text{III}}$ shift towards zero, until for FM $|J_1|\!\ge\!4J_3$ the
two minima merge at \qv\,=\,0 which becomes the global minimum.

These additional minima at incommensurate $\qv_{\text{II}}$ and
$\qv_{\text{III}}$ imply possible formation of helical magnetic order, but the
exact picture does depend on interlayer exchange coupling $J_z$ too,
\cite{DMH+11,CRY+11} the latter being one of possible way to stabilize the
magnetic structures observed in the ``intermediate'' systems $M$CrS$_2$
($M$=Cu, Au, Ag, Na).  In particular, this may be the origin of incommensurate
magnetic structures for $M$=Cu, Na, or commensurate double stripes for $M$=Ag,
Au.

An extra complication is introduced by the observed monoclinic distortion in
AgCrS$_2$ and AuCrS$_2$, which induces three pairs of nonequivalent
nearest-neighbor exchange couplings ($J_{1x}$, $J_{1y}$), ($J_{2x}$,
$J_{2y}$), and ($J_{3x}$, $J_{3y}$) (see Fig.~\ref{fig:90_ds}).  The observed
four-sublattice spin arrangement cancels the effect of $J_{1x}$ and
$J_{2y}$. In order to clarify which of remaining magnetic exchanges are
relevant for the stabilization of the DS structure, namely the ferromagnetic
first-neighbor coupling $J_{1y}$, the antiferromagnetic second-neighbor
$J_{2x}$ and antiferromagnetic third-neighbors $J_{3x}$ and $J_{3y}$
superexchanges, and the interplane antiferromagnetic superexchange $J_z$ we
calculated the energy of different Cr spin moment configurations and derived
the corresponding exchange values. The results are summarized in Table
\ref{table:LTJs}. They show that the monoclinic distortion does stabilize the
DS structure by strongly suppressing the AFM contribution to $J_{1y}$ along
the FM Cr chains.

\begin{table}[tbh]
\caption{\label{table:LTJs} LSDA exchange coupling constants (in meV) for low
temperature phases of AgCrS$_2$ and AuCrS$_2$.}
\begin{ruledtabular}
\begin{tabular}{c........}
  $M$ & \multicolumn{1}{c}{$J_{1x}$} & \multicolumn{1}{c}{$J_{1y}$}
  & \multicolumn{1}{c}{$J_{2x}$} & \multicolumn{1}{c}{$J_{2y}$}
  & \multicolumn{1}{c}{$J_{3x}$} & \multicolumn{1}{c}{$J_{3y}$}
  & \multicolumn{1}{c}{$J_{zx}$} & \multicolumn{1}{c}{$J_{zy}$} \\
 Au & 5.14 & 2.69 & 0.70 & 0.57 & 3.19 & 3.24 & 1.85 & 1.32 \\
 Ag & 1.12 & -1.36 & -0.30 & -0.23 & 2.54 & 2.62 & 1.09 & 0.62 \\
\end{tabular}
\end{ruledtabular}
\end{table}

We also have to comment on the values of exchange constants for AuCrS$_2$
shown in Tables~\ref{table:HTJs_LSDA} and \ref{table:LTJs}. These values
definitely deviate from the regularities observed in other materials of this
series.  The ratio of the important exchange constants J$_3$ and J$_1$ for
AuCrS$_2$ is still such that it gives the double-stripe structure observed
experimentally.  However the absolute values of these exchanges for this
system are about two times larger than what one would expect from the
comparison with other materials of this class. We do not have a full
explanation of this difference. A possible reason is that AuCrS$_2$ has a
delafossite structure with interlayer Au$^+$ ions in a linear
coordination.\cite{Pabst-46} It is possible that the reason for different
values of exchange for this system is connected with that. Still, this
situation is definitely unsatisfactory, and it requires further study.

\subsection{\label{sec:magprop}Interpretation of magnetic properties}

Our calculations, presented above, have shown that indeed the observed types
of magnetic ordering in Cr-plane in $M$CrS$_2$ ($120^{\circ}$ for Li, double
stripes for Ag and Au, ferro layers for K) are reproduced. The obtained values
of exchange constants, Table~\ref{table:HTJs_LSDA}, allow to explain these
magnetic structures.

Thus for the smallest $M$-ion Li the nn exchange $J_1$ is the strongest and
antiferromagnetic; apparently it is predominantly responsible for the observed
pure antiferromagnetic ($120^{\circ}$) ordering, observed in LiCrS$_2$. With
the increasing of Cr-Cr distance and Cr-S-Cr angle
(Li$\rightarrow$Cu$\rightarrow$Au$\rightarrow$Ag$\rightarrow$Na$\rightarrow$K)
the value of $J_1$ decreases and then changes sign, becoming ferromagnetic for
$M$=Ag, K. Simultaneously the AF exchanges between 3-d neighbors $J_3$ remains
relatively large, and it plays important role for intermediate compounds
Ag/AuCrS$_2$, apparently leading to their double-stripe ordering. Finally,
large nearest neighbor ferro interaction $J_1$ for the large M-ion K
guarantees ferro ordering in Cr-plane in KCrS$_2$.

\begin{figure}[tbp!]
\includegraphics[width=0.47\textwidth]{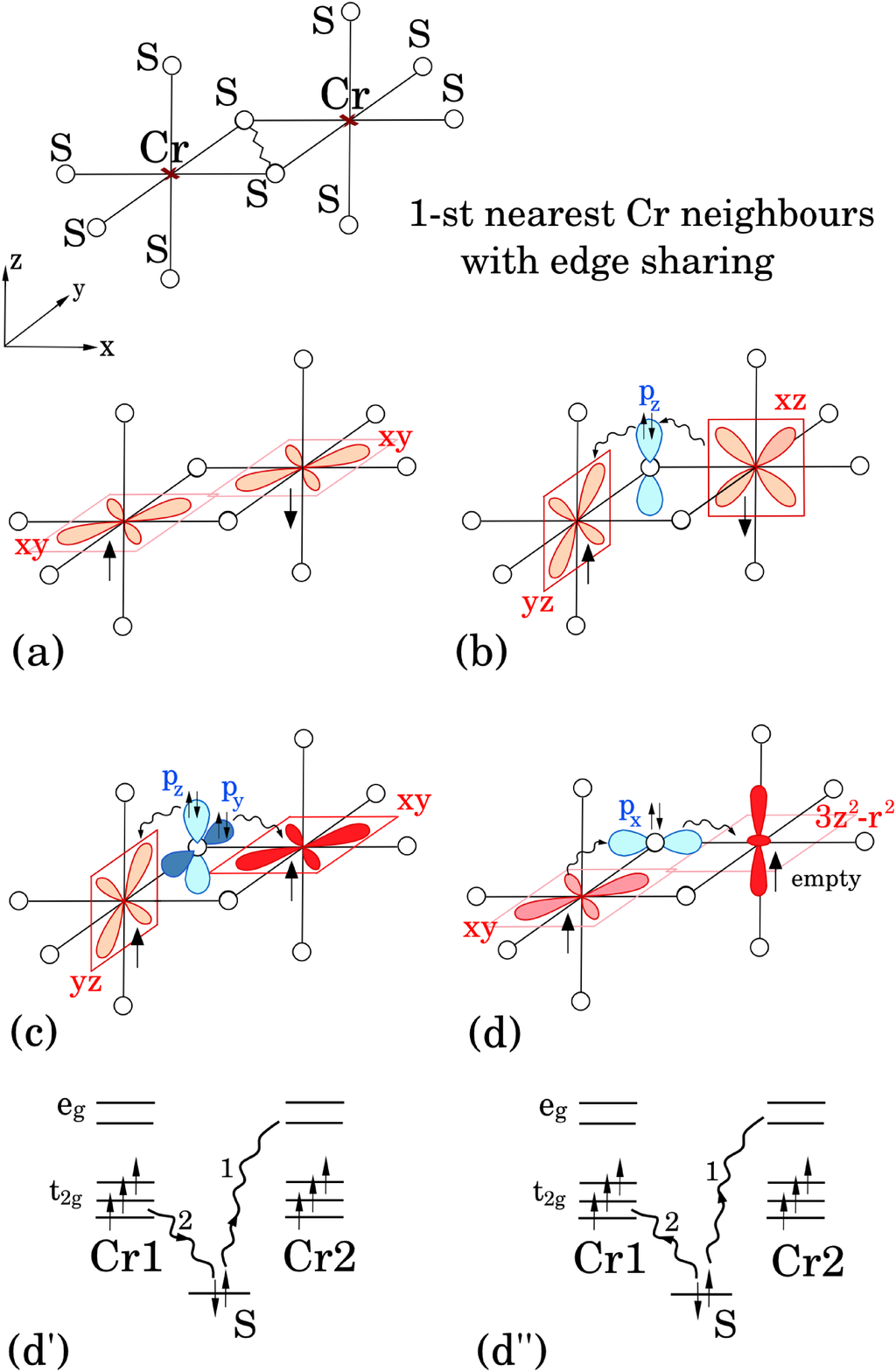}
\caption{\label{fig:J1s}(Color online) All possible contributions to $J_1$}
\end{figure}

To understand the microscopic origin of different exchange integrals in this
series, one should look at different microscope exchange passes.  In
Fig.~\ref{fig:J1s}(a)-\ref{fig:J1s}(d) we show the main paths of
superexchange, existing in CrS$_2$ planes with the Cr$^{3+}$ ions with
$d$-shell t$_{2g}^3$e$_g^0$ and with the geometry of edge-sharing CrS$_6$
octahedra with nearest neighbor Cr-S-Cr angle of about 90$^{\circ}$.

First of all, there exist a direct overlap of different {\ttg} orbitals of
neighboring Cr ions, e.g., $xy$-orbital in Fig.~\ref{fig:J1s}(a). It gives a
rather large AF exchange
\begin{equation}
\label{eq:ja}
J_{a} \sim \frac{t_{dd}^2}{U_{dd}},
\end{equation}
which however strongly decreases with the increasing of Cr-Cr distance.

In Fig.~\ref{fig:J1s}(b) and \ref{fig:J1s}(c) we show an exchange of
\ttg-\ttg\ via 90$^{\circ}$ Cr-S-Cr bond. The process~\ref{fig:J1s}(b)
(virtual hopping of {\ttg} electrons through the same ligand $p$-orbital, in
this case $p_z$), gives strong AF exchange:
\begin{eqnarray}
J_{b} \sim
\frac{t_{pd\pi}^4}{\Delta^2}\left(\frac{1}{\Delta}+\frac{1}{U_{dd}}\right),
\label{eq:jb}
\end{eqnarray}
where we denoted by $\Delta$ the charge-transfer energy (the energy of a
transition Cr$^{3+}$($d^3$)S$^{2-}$($3p^6$) $\rightarrow$
Cr$^{2+}$($d^4$)S$^{-}$($3p^5$)). One sees that this process does not change
strongly with the Cr-S-Cr angle, only the distance Cr-S determines the value
of $t_{pd\pi}$ hopping, and this distance is approximately constant in the
whole series $M$CrS$_2$.

The process~\ref{fig:J1s}(c) (the \ttg-\ttg\ exchanges via different
S\,$p$-orbitals) leads to the ferromagnetic exchange, which is however usually
weaker,
\begin{eqnarray}
\label{eq:jc}
J_{c} \sim - \frac{t_{pd\pi}^4}{\Delta^3} \times \frac{9J_{H,S}}{\Delta},
\end{eqnarray}
(here $J_{H,S}$ is the Hund's rule coupling on sulfur) and it decreases by
absolute value with decreasing Cr-S-Cr angle.

More important is another ferromagnetic contribution due to a virtual hopping
from the occupied {\ttg}-shell of one Cr to the empty {\eg}-shells of another
Cr, Fig.~\ref{fig:J1s}(d).  As one sees from Fig.~\ref{fig:J1s}(d) this
process also gives ferromagnetic contribution,
\begin{eqnarray}
\label{eq:jd}
J_{d}&\sim&
-\frac{t_{pd\sigma}^2t_{pd\pi}^2}{\Delta^2U_{dd}}\times\frac{3J_H}{U_{dd}} \nn \\
&&- \frac{t_{pd\sigma}^2t_{pd\pi}^2}{\Delta^3}\times
\frac{3J_H}{\Delta},
\end{eqnarray}
where the first term corresponds to a process~\ref{fig:J1s}(d$'$) (effective
transfer of an electron from one Cr to the other via S), and the
second~\ref{fig:J1s}(d$''$) (the transfer of two $3p$-electrons of S to the
left and right Cr ions).  We do not keep here some numerical coefficients.
Note that despite the presence of a small factors $\frac{J_H}{U_{dd}}$ or
$\frac{J_H}{\Delta}$, this ferromagnetic contribution~(\ref{eq:jd}) is
comparable with~(\ref{eq:ja}) (typically $t_{pd\sigma}\sim
\sqrt{2}t_{pd\pi}$), and also the Hund's rule contribution in~(\ref{eq:jd}) is
enhanced by factor 3.  Thus, though usually the 90$^{\circ}$-exchange
involving Hund's rule interaction gives ferromagnetic, but weaker exchange, in
this case due to a specific electrons occupation of Cr$^{3+}$ it can give
significant contribution and can even start to dominate if the other
competitive contributions are small. This is apparently what happens in
KCrS$_2$, in which the main competing AF exchange~\ref{fig:J1s}(a) is strongly
reduced due to a large size of K$^{+}$ and corresponding increase of Cr-Cr
distance.

\begin{figure}[tbp!]
\includegraphics[width=0.47\textwidth]{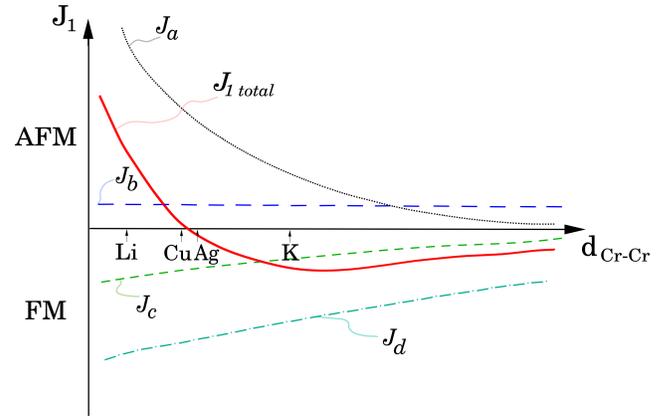}
\caption{\label{fig:expl}(Color online) Schematic dependence of different
contributions to the nearest neighbor Cr-Cr exchange $J_1$.}
\end{figure}

Thus, we can schematically present different contributions to the nearest
neighbor exchange $J_1$ and their change in the row
(Li$\rightarrow$Cu$\rightarrow$Au$\rightarrow$Ag$\rightarrow$Na$\rightarrow$K)CrS$_2$
as following, Fig.~\ref{fig:expl}.

\begin{figure}[tbp!]
\includegraphics[width=0.45\textwidth]{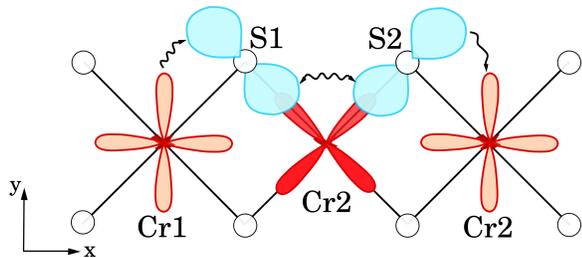}
\caption{\label{fig:J3s}(Color online) A possible exchange path
contributing to the antiferromagnetic exchange of third neighbors
$J_3$.}
\end{figure}

To explain the resulting magnetic structures, especially double-stripe
structure of AuCrS$_2$ and AgCrS$_2$, we also have to include the further
neighbor exchange. As seen from Table~\ref{table:HTJs_LSDA}, the second
neighbor exchange is always small. Somewhat surprisingly, larger and more
important turns out to be the interaction of third neighbors. It can be
schematically explained by the consideration shown in Fig.~\ref{fig:J3s}, in
which one sees that there is an exchange path connecting occupied $t_{2g}$
orbitals on third neighbors Cr$_1$ and Cr$_3$ via two sulfur S$_1$ and S$_2$,
(with their $p$-orbitals being relatively large) due to the $p$-$p$ overlap,
or to the overlap via an empty {\eg}-orbital ($x^2-y^2$) of Cr$_2$
(Fig.~\ref{fig:J3s}).

Thus in this geometry the coupling between third neighbors $J_3$ turns out to
be reasonably large (larger than $J_2$) and antiferromagnetic, and in effect
it is this coupling which stabilizes double-stripe structure for
``intermediate'' composition AgCrS$_2$ and AuCrS$_2$, in which the main
nearest neighbor interaction $J_1$ is small due to a compensation of different
contributions to it.

The general tendency showing regular change of different exchange
contributions, especially of the nearest neighbor exchange $J_1$, see
Fig.~\ref{fig:expl}, is also confirmed by model calculations in which we took
LiCrS$_2$ and artificially compressed it in $c$-direction, keeping the volume
constant. At this change the in-plane Cr-Cr distance and Cr-S-Cr angle
increase, following the same trends as in going from LiCrS$_2$ to (Ag, Au) and
to KCrS$_2$.  Our \textit{ab-initio} calculations of this model system
confirmed the trend discussed above: with increasing Cr-Cr distance large AF
coupling $J_1$ strongly decreases and becomes ferromagnetic.

\subsection{\label{sec:ldau} The effect of LSDA+$U$ on calculated exchange
constants}

So far we discussed only exchange coupling constants determined by fitting
$E(\qv)$ curves calculated within LSDA (Table
\ref{table:HTJs_LSDA}). Comparing the $J_1/J_3$ ratio from Table
\ref{table:HTJs_LSDA} with the critical values obtained from the analysis of
the $J_1$--$J_3$ Heisenberg model one notices that for some of the compounds
the estimated $J_i$ do not give an experimentally observed ground state. For
KCrS$_2$, for example, $J_1/J_3=-2.6>-4$ corresponds to an incommensurate
spin-spiral structure in the $a$-$b$ plane instead of experimental FM
ordering.

One of possible reasons for this is that the LSDA underestimates the Coulomb
repulsion between rather localized Cr\,3$d$ electrons. The $U_{dd}$ parameter
in expressions (\ref{eq:ja})--(\ref{eq:jd}) is the energy cost of adding an
electron to one of the unoccupied minority-spin \ttg\ state. In LSDA it is
governed solely by the exchange splitting of about 2.4 eV between the
minority- and majority-spin \ttg\ states, i.e., by Hund's coupling of $3J_H$.
As a result the LSDA overestimates those contributions to the inter-site
exchanges that have $U_{dd}$ in the denominator. Accounting for the Coulomb
repulsion in LSDA+$U$ calculations increases the energy difference between the
minority- and majority-spin Cr \ttg\ states by $U_{\text{eff}}$, so that
$U_{dd}$ becomes equal $3J_H+U_{\text{eff}}$.

The increase of $U_{\text{eff}}$ suppresses AFM $J_{a}$ and $J_{b}$, whereas
the FM \ttg--\eg\ contribution $J_d$ is much less affected. Thus, in the
compounds with FM $J_1$ ($M=$\,Na, K) it becomes even stronger, whereas in
those compounds for which LSDA gives AFM $J_1$ its value decreases and it may
even change sign. On the other hand, the AFM 3-rd neighbor coupling $J_3$,
which is governed by the \ttg--\ttg\ superexchange (Sec.\ \ref{sec:magprop}),
gradually decreases with the increase of $U_{\text{eff}}$.

This combined effect of strengthening the FM $J_1$ and weakening the AFM $J_3$
leads to a reduction in the $J_1/J_3$ ratio estimated for KCrS$_2$ from $-$2.6
in LSDA to $-$4.6 and $-$6.4 in LSDA+$U$ calculations with $U_{\text{eff}}=1$
and 2 eV, respectively. Thus, accounting for Coulomb repulsion stabilizes the
FM in-plane order in KCrS$_2$. In LiCrS$_2$ the 120$^{\circ}$ structure gives
the lowest total energy also in LSDA+$U$ calculations. In other compounds the
increase of $U_{\text{eff}}$ changes the $J_1/J_3$ ratio and, consequently,
the position of incommensurate minima.

\section{\label{sec:summ}Summary}

Summarizing, the results of our \textit{ab-initio} calculations, and model
considerations of Sec.~\ref{sec:magprop} allowed us to explain the very
interesting sequence of magnetic phases in layered chromites $M$CrS$_2$ with
triangular Cr layers, in which the magnetic ordering in Cr layers changes from
purely antiferromagnetic (120$^{\circ}$) structure in LiCrO$_2$ via
``intermediate'' double-stripe structure of AgCrS$_2$ and AuCrS$_2$ (and
incommensurate structure in NaCrS$_2$ and CuCrS$_2$) to a ferromagnetic layers
in KCrS$_2$. These structures emerge mainly as a result of competing
contributions to the nearest neighbor exchange $J_1$, together with reasonably
large antiferromagnetic exchange for third neighbors $J_3$. In particular,
their combined action leads to the most interesting double stripe structure of
AuCrS$_2$ and AgCrS$_2$, which apparently is responsible for the multiferroic
behavior of the latter (and probably also in the former --- it is not checked
yet). Our study demonstrates quite nontrivial interplay of lattice geometry
and orbital occupation in giving such diverse magnetic behavior in apparently
rather similar materials. The frustrated nature of the lattice definitely
plays a very important role in these phenomena. Such high sensitivity of
magnetic, and apparently some other, e.g., multiferroic properties to fine
details of electronic and lattice structure could probably be used also to
tune the properties of other similar materials. We envisage that further
studies of the stability of nuclear and magnetic structures may provide a clue
to tailor the magnetoelastic coupling and the multiferroic properties in
geometrically frustrated oxides, sulfides and selenides with different
transition metals.

\section*{Acknowledgments}

A.~V.~Ushakov and D.~A.~Kukusta gratefully acknowledge the hospitality at
Max-Planck-Institut f\"ur Festk\"orperforschung in Stuttgart during their stay
there.

The authors are grateful to S.~Hebert and C.~Martin for discussion of
experimental situation. The work of A.~U.\ and D.~Kh.\ was supported by the
European program SOPRANO and by the German projects SFB 608 and FOR 1346.

%

\end{document}